\documentclass[aps,prb,twocolumn,showpacs,nofootinbib,superscriptaddress]{revtex4-1}
\usepackage{graphicx}
\usepackage{amsmath,latexsym}
\usepackage{amssymb}
\usepackage{dcolumn}
\usepackage{bm}
\usepackage{xcolor}
\usepackage{braket}
\newcommand{\nn}{\nonumber}
\newcommand{\beq}{\begin{eqnarray}}
\newcommand{\eeq}{\end{eqnarray}}

\begin{document}
\title{Charge/Spin Supercurrent and the Fulde-Ferrell State Induced by Crystal Deformation in Weyl/Dirac Superconductors}

\author{Taiki Matsushita}
\affiliation{Department of Materials Engineering Science, Osaka University, Toyonaka, Osaka 560-8531, Japan}
\author{Tianyu Liu}
\affiliation{Department of Materials Engineering Science, Osaka University, Toyonaka, Osaka 560-8531, Japan}
\affiliation{Department of Physics and Astronomy, University of British Columbia, Vancouver, BC, Canada V6T 1Z1}
\affiliation{Quantum Matter Institute, University of British Columbia, Vancouver BC, Canada V6T 1Z4}
\author{Takeshi Mizushima }
\affiliation{Department of Materials Engineering Science, Osaka University, Toyonaka, Osaka 560-8531, Japan}
\author{Satoshi Fujimoto}
\affiliation{Department of Materials Engineering Science, Osaka University, Toyonaka, Osaka 560-8531, Japan}

\date{\today}
\begin{abstract}
It has been predicted that emergent chiral magnetic fields can be generated by crystal deformation in Weyl/Dirac metals and superconductors. The emergent fields give rise to chiral anomaly phenomena as in the case of Weyl semimetals with usual electromagnetic fields. Here, we clarify effects of the chiral magnetic field on Cooper pairs in Weyl/Dirac superconductors on the basis of the Ginzburg-Landau equation
microscopically derived from the quasiclassical Eilenberger formalism. It is found that Cooper pairs are affected by the emergent chiral magnetic field in a dramatic way, and the pseudo-Lorentz force due to the chiral magnetic field stabilizes the Fulde-Ferrell state and causes a charge/spin supercurrent which flows parallel to the chiral magnetic field in the case of Weyl/Dirac superconductors. This effect is in analogy with the chiral magnetic effect (CME) of Weyl semimetals. In addition, we elucidate that neither Meissner effect nor vortex state due to chiral magnetic fields occurs. 
\end{abstract}
\maketitle

\section{\label{sec:level1}Introduction}
Surface states of topological insulators and topological superconductors are endowed with Dirac and Majorana fermions as low-energy excitations protected by topology. \cite{Kane1,Kane2,hasan-kane,Xiao topo} The notion of the topologically protected gapless fermionic excitations has been extended to bulk metallic systems, i.e. Dirac/Weyl semimetals. \cite{mura,Wan,burk,Xu,bale,kim,wang,wang2,neup,boris,liujiang,xu2,lv,schekhar,huang} 
Both the Dirac semimetals (DSMs) and Weyl semimetals (WSMs) are characterized by the monopole charges of the Dirac/Weyl points in the momentum space. In the former systems, the monopole charges with opposite signs are located at the same point, and  hence, the protection mechanism based on symmetry, e.g.  crystal rotational symmetry, is necessary for their stability.
On the other hand,  in the latter systems, the monopoles with opposite signs are separated in the momentum or energy space 
because of broken time-reversal symmetry or inversion symmetry, and thus they are stable against any symmetry-breaking perturbations.  
In WSMs,  the monopoles in the momentum space give rise to various intriguing transport phenomena associated with chiral anomaly, such as the anomalous Hall effect, 
chiral magnetic effect, and negative magnetoresistivity. \cite{Xu,kim,huang,ABJ anomaly,fuku,Son,Yu,zyuzin,gos,gos2}

Recently, it has been revealed that crystal deformation affects low energy properties of Weyl and Dirac metals in a dramatic way, and generates an emergent U(1) chiral gauge field.\cite{Hughes, Onker, hughes-ryu,Qi,franz,sumiyoshi,grushin,zubkov,cort,you,Liu}
The idea of generating an emergent electromagnetic field from lattice distortion was originally proposed and examined for graphene, which is a classical Dirac fermion system
in condensed matter physics. \cite{Guinea,Levy}  Crystal deformation yields the nontrivial spatial dependence of the position of Dirac points, which plays a role of a fictitious vector potential, leading to an emergent magnetic field. 
This idea can be naturally generalized to cases of DSMs and WSMs. It has been predicted that a uniform emergent chiral magnetic field can be realized by applying strain in DSMs,
which leads to the Landau quantization of the energy spectrum. \cite{franz,Liu} 
The theory is applicable to real Dirac materials such as Cd$_3$As$_2$ and Nb$_3$Pb.
Also, the emergent chiral electromagnetic field can give rise to  chiral anomaly phenomena in WSMs, as in the case with usual electromagnetic fields. \cite{Hughes, Onker, hughes-ryu,franz,sumiyoshi,grushin}
For instance, lattice defects such as dislocation also give rise to emergent chiral magnetic fields, which lead to the torsional chiral magnetic effect (or chiral torsional effect) in WSMs with broken time-reversal symmetry; i.e. 
an equilibrium current flowing along dislocation lines is generated without applying real electromagnetic fields.\cite{sumiyoshi}
More generally, various spatially inhomogeneous structures of crystal lattices and order parameters such as spontaneous magnetization
can be sources of emergent chiral electromagnetic fields acting on Dirac/Weyl quasiparticles.

In this paper, we investigate effects of strain-induced chiral magnetic fields in Weyl superconductors (WSCs) and Dirac superconductors (DSCs). \cite{bale-weyl,yang-dirac,TSC}  
It is noted that the coupling charge of chiral magnetic fields of electrons and holes have the same sign, and hence, 
the Meissner effect of chiral magnetic fields due to supercurrents does not occur.
Because of this feature, the chiral magnetic fields lead to 
the Landau quantization of the energy spectrum  of Dirac/Weyl quasiparticles. \cite{Liu2} 
Furthermore, chiral anomaly phenomena arising from spatially inhomogeneous structures of the superconducting order parameter
in the Weyl superconducting state have been extensively studied so far, particularly, for the ABM phase of the superfluid Helium 3. 
\cite{volovik1,volovik2} 
However, effects of chiral magnetic fields on Cooper pairs composed of Weyl/Dirac quasiparticles are not well understood so far.
The purpose of this paper is to address this issue 
on the basis of a microscopically derived Ginzburg-Landau (GL) equation.
As mentioned above, the emergent chiral magnetic field does not give rise to the Meissner effect.
Nevertheless, it still affects dynamics of Cooper pairs via an interaction with a pseudo-Lorentz force generated by the chiral magnetic field.
It is well-known that the usual Lorentz force proportional to $\vec{v} \times \vec{B}$, where $\vec{v}$ is the velocity and $\vec{B}$ is a magnetic field, does not directly couple to Cooper pairs, because Cooper pairs consist of electrons with momentum $\vec{k}$ and $-\vec{k}$, and thus, the Lorentz force acting on these two electrons cancels with each other.\cite{Houghton}
In contrast, the pseudo-Lorentz force indeed interacts with Cooper pairs, since the coupling charge of strain-induced 
chiral magnetic fields depends on chirality, and two electrons which constitute a Cooper pair carry opposite chiralities as well
as opposite signs of momenta.

Here we clarify some nontrivial effects arising from the pseudo-Lorentz force acting on Cooper pairs.
The main results are as follows.
An emergent chiral magnetic field $\vec{B}^{\rm em}$ induced by lattice distortion gives rise to supercurrents $\vec{J}_{\rm s}$ flowing parallel to the direction of the chiral magnetic field
in WSCs:
\begin{eqnarray}
\vec{J}_{\rm s}=\alpha _{\rm W}\vec{B}^{\rm em},
\end{eqnarray}
where $\alpha_{\rm W}$ is a constant determined from material parameters.
 It is noted that this is not a Meissner current, the direction of which should be perpendicular to applied magnetic fields.
The supercurrent parallel to the chiral magnetic field is akin to the chiral magnetic effect of Weyl semimetals, i.e.
a charge current induced by a magnetic field.\cite{fuku}
However, we stress that this effect is not directly related to chiral anomaly of Weyl fermions, but caused by
the coupling of Cooper pairs with the pseudo-Lorentz force.
Accordingly, the Fulde-Ferrell state, in which the phase of the superconducting gap is spatially inhomogeneous,
is stabilized in the case that bulk supercurrent flow is prohibited because of boundary conditions.\cite{ff}

In the case of DSCs, where time-reversal symmetry is preserved, 
assuming a spin-triplet pairing state, we elucidate that 
the chiral magnetic field due to lattice deformation generates a spin supercurrent,
\begin{eqnarray}
\vec{J}_{\rm s}^{\rm spin}=\alpha _{\rm D}\vec{B}^{\rm em},
\end{eqnarray}
where $\alpha_{\rm D}$  is a response coefficient.
This result also implies that the realization of the spin-dependent Fulde-Ferrell state, where
Cooper pairs with opposite spin directions has opposite signs of the phase of the superconducting gap.
These findings for DSCs is relevant to Cd$_3$As$_2$, which shows superconductivity under applied pressure. \cite{L.P.He}
It  has been argued that 
momentum-orbital locking in the DSM suppresses ordinary spin-singlet $s$-wave pairings, and hence, a spin-triplet orbital-singlet pairing state is realized.\cite{TSC Dirac,SC Dirac} 
In this pairing state, the Dirac points in the normal state result in the existence of point-nodes of the superconducting gap, characterizing the Dirac superconducting state.
As long as the chemical potential is located close to the Dirac points, the emergent chiral magnetic field acts on
Cooper pairs which are composed of electrons on the Fermi surfaces surrounding the Dirac points.

The organization of this paper is as follows.
In Sec. \ref{sec:level2}, we consider the case of WSCs using a simple toy model. 
In this toy model, we introduce a chiral magnetic field by hand without referring to its physical origin.
Although the model is not directly related to real materials, the analysis of this toy model is useful
for qualitative understanding  of effects of the pseudo-Lorentz force on Cooper pairs.
Exploiting the quasiclassical Eilenberger equations, we derive microscopically the GL equation,
in which the pseudo-Lorentz force is incorporated. 
It is found that although the chiral magnetic field does not lead to the Meissner effect, it gives rise to
supercurrents flowing parallel to the chiral magnetic field, and that the Fulde-Ferrell state is realized in the
case that a bulk supercurrent flow is prohibited by boundary conditions. 
 In Sec. \ref{sec:level4}, the case of DSCs is investigated on the basis of the effective two-orbital model for DSMs in which  
 Dirac points are protected by
 $C_{4}$ rotational symmetry. This model is relevant to Cd$_3$As$_2$ and Nb$_3$Pb.
 Lattice strain which gives rise to a chiral magnetic field is explicitly incorporated in this two-orbital model.
The analysis based on this microscopic model confirms  the predictions obtained for the toy model in Sec. \ref{sec:level2}.
Assuming a spin-triplet orbital-singlet pairing state,\cite{SC Dirac} and Exploiting the Eilenberger equations, 
we demonstrate that lattice torsion induces spin supercurrent flow.
This result also implies that in the situation that the bulk current flow is prohibited by boundary conditions,
the spin-dependent Fulde-Ferrell state is realized.
Also, we consider an effect of the Zeeman magnetic field applied to the system, which generates a charge supercurrent
from a spin supercurrent. Summary is given in Sec. \ref{sec:level5}.


\section{\label{sec:level2}Weyl superconductor with emergent chiral magnetic field}

In this section, we consider 
a three-dimensional chiral $p$-wave superconductor 
with the gap function $\Delta_{\bm{k}}=\Delta(k_x+ik_y)$,
which is a typical example of a Weyl superconductor. To incorporate an emergent chiral magnetic field due to lattice distortion into our system, we need to deal with a multi-orbital model. However, it is desirable to avoid extrinsic complexity of the analysis arising from the multi-orbital character. To focus on qualitative understanding of effects of chiral magnetic fields on Cooper pairs, we, here, exploit a toy model in which a chiral vector potential is incorporated by hand without referring to its microscopic origin. 

In the following, we, first, derive the quasiclassical Eilenberger equation. From this equation, it is found that the chiral vector potential gives rise to the pseudo-Lorentz force acting on Cooper pairs as well as Bogoliubov quasiparticles. We elucidate that this effect results in  supercurrent flow parallel to the chiral magnetic field, which is akin to the chiral magnetic effect of Weyl semimetals. Furthermore, it is found that the Fulde-Ferrell state, i.e. an inhomogeneous superconducting state with a spatially-varying phase, is realized, when the bulk current flow is prohibited by boundary conditions.

\subsection{\label{sec:2-A}Eilenberger equation for a $p+ip$ superconductor with chiral magnetic fields}
 
We consider a toy model of a spinless chiral $p$-wave superconductor with a chiral vector potential. This toy model is used for qualitative understanding of effects of chiral magnetic fields on Cooper pair dynamics. The chiral $p+ip$ pairing state has the point nodes of the superconducting gap at north and south poles on the Fermi sphere, which accompany monopole charges in the momentum space. In the vicinity of the point nodes, Bogoliubov quasiparticles behave as Weyl fermions. \cite{bale-weyl} 
The Hamiltonian for the three-dimensional spinless chiral $p$-wave superconductor coupled with both a usual vector potential  and a chiral vector potential is given by, 
\begin{widetext}
\begin{eqnarray}
H&=&\int d^3x 
 \psi^{\dag} (\vec{x}) \left[ \frac{\left(-i\nabla_x-e\vec{A}(\vec{x})-e\vec{A}^{\rm em}(\vec{x},\eta)\right)^2-k_F^2}{2m}  \right] \psi(\vec{x})\nn\\&+&\int d^3x \left[\psi^{\dag} (\vec{x})\left\{\frac{\Delta}{2k_F},\left(-i\frac{\partial}{\partial x}-\frac{\partial}{\partial y} \right)\right\}\psi^{\dag} (\vec{x})+\psi (\vec{x})\left\{\frac{\Delta^*}{2k_F},\left(-i\frac{\partial}{\partial x}+\frac{\partial}{\partial y} \right)\right\}\psi (\vec{x})\right],
\end{eqnarray}
\end{widetext}
where $\psi$ and $\psi^\dag$ are annihilation and creation operators of spinless fermions with mass $m$ and charge $e$, $\Delta$ is the superconducting gap, $k_F$ is the Fermi momentum, and $\{a, b\}=ab+ba$. $\vec{A}(\vec{x})$ is a vector potential for a real magnetic field, and $\vec{A}^{\rm em}(\vec{x},\eta)$ is  an emergent chiral vector potential which is originated from crystal deformation. $\eta=\pm1$ is chirality of a Weyl point in the momentum space. It is noted that $\vec{A}^{\rm em}$ depends on the chirality of Weyl point, and satisfies $\vec{A}^{\rm em}(\vec{x},-\eta)=-\vec{A}^{\rm em}(\vec{x},\eta)$, because of its chiral character. \cite{Hughes,Onker}
In the following, 
we discuss unusual transport phenomena induced by the emergent chiral vector potential using the quasi-classical Eilenberger equation based on the Matsubara Green's function formalism in the Nambu-Gor'kov space. 
We assume clean limit for simplicity. The left-hand and right-hand Gor'kov equations for this system are given by
\begin{eqnarray}
[ -i\omega_n\check{\tau_3}+\check{H}_{{\rm Nambu}}(\vec{x},-i\nabla_x) ] \check{G}(\vec{x},\vec{x}^{\prime},\omega_n)\nn\\=\delta(x-x^{\prime})\check{1}, \label{eq:gor1} \\
\check{G}(\vec{x},\vec{x}^{\prime},\omega_n)[ -i\omega_n\check{\tau_3}+\check{H}_{{\rm Nambu}}(\vec{x'},-i\nabla_{x'}) ]\nn\\=\delta(x-x^{\prime})\check{1}, \label{eq:gor2}
\end{eqnarray}
where $\omega_n=(2n+1)\pi T$ is the Matsubara frequency. $\check{\tau}_j\; (j=1,2,3)$ and $\check{1}$ are, respectively, the Pauli matrix and identity matrix in the Nambu-Gor'kov space. Here, we introduced a matrix representation of the Hamiltonian in the Nambu-Gor'kov space,
\begin{widetext}
\begin{eqnarray}
\check{H}_{{\rm Nambu}}(\vec{x},-i\nabla_{x})=\check{\zeta}(\vec{x},-i\nabla_{x})+\check{\Delta}(\vec{x},-i\nabla_{x}),
\end{eqnarray}
\begin{eqnarray}
\check{\zeta}(\vec{x},-i\nabla_{x})
=\begin{pmatrix}
\frac{[-i\nabla_{x}-e\vec{A}(\vec{x})-e\vec{A}^{\rm em}(\vec{x},\eta)]^2-k_F^2}{2m}&&0\\
0&&\frac{[i\nabla_{x}-e\vec{A}(\vec{x})+e\vec{A}^{\rm em}(\vec{x},\eta)]^2-k_F^2}{2m}
\end{pmatrix},
\end{eqnarray}
\begin{eqnarray}
\check{\Delta}&=&\begin{pmatrix}
0&&-\left\{\frac{\Delta}{2k_F},(-i\frac{\partial}{\partial x}-\frac{\partial}{\partial y})\right\}\\
\left\{\frac{\Delta^*}{2k_F},(-i\frac{\partial}{\partial x}+\frac{\partial}{\partial y})\right\}&&0
\end{pmatrix},
\end{eqnarray}
\end{widetext}
where $\check{\zeta}$ and $\check{\Delta}$ are, respectively, the kinetic energy part and the gap function part of the Hamiltonian.
The Matsubara Green's function $\check{G}(x,x')$ in the superconducting state is defined as, 
\begin{eqnarray} 
\label{eq:green1}
\check{G}=
\begin{pmatrix}
G&F\\-F^\dag&\overline{G}
\end{pmatrix},
\end{eqnarray}
\begin{gather}
 \label{eq:greenSC}
G(x_1,x_2)\equiv\langle T_\tau\psi(x_1)\psi^\dag(x_2) \rangle,\\
F^\dag(x_1,x_2)\equiv\langle T_\tau\psi^\dag(x_1)\psi^\dag(x_2) \rangle,
\end{gather}
\begin{gather}
F_(x_1,x_2)\equiv\langle T_\tau\psi(x_1)\psi(x_2) \rangle,\\
\overline{G}(x_1,x_2)\equiv-\langle T_\tau\psi^\dag(x_1)\psi(x_2)\rangle,
\end{gather}
where $\psi(x)$ and $\psi^\dag(x)$ are annihilation and creation operators with $x=(\tau,\vec{x})$. 
$T_{\tau}$ is the $T$-ordering operator for the imaginary time $\tau$.

We now apply quasiclassical approximation to this model, which is valid for $k_F\xi \gg 1$ with $\xi$ the coherence length, and derive the Eilenberger equations for the quasiclassical Green's function by performing energy integral of the Gor'kov equations. \cite{Eilenberger} The quasiclassical Green's function is defined as follow,
\begin{align}
\label{quagreen}
\check{g}(\vec{k_{\parallel}},\vec{R},\omega_n)\equiv \int{}\frac{d\zeta}{i\pi}\check{G}(\vec{k},\vec{R},\omega_n)=\begin{pmatrix}
g&&f\\
-f^\dag&&\overline{g}
\end{pmatrix},
\end{align} 
where $k_\parallel$ is the momentum component parallel to the Fermi surface, $\vec{R}=(\vec{x}_1+\vec{x}_2)/2$ is the center of mass coordinate of a Cooper pair, $\zeta=(k^2-k^2_F)/(2m)$, 
and $\check{G}(\vec{k},\vec{R},\omega_n)$ is the Fourier transform of Green's function $\check{G}(\vec{x}_1,\vec{x}_2,\omega_n)$ with momentum $\vec{k}$ reciprocal to $\vec{x}_1-\vec{x}_2$. Following the standard procedure,\cite{Eilenberger,Houghton}  we derive the Eilenberger equation satisfied by the quasiclassical Green's function up to the leading order in $1/(k_F\xi) \ll 1$,
\begin{widetext}
\begin{align}
\label{2-15}
[i\omega_n\check{\tau}_3-\check{\Delta}+e\vec{v}_{\rm F}\cdot\vec{A}\check{\tau}_3, \check{g}]
+i\left(\vec{v}_F-\frac{e}{m}\vec{A}^{\rm em}\right)\cdot\nabla_R\check{g}
+
\frac{ie}{2}\vec{v}_{\rm F}\times\vec{B}\cdot\frac{\partial}{\partial\vec{k}_{\parallel}}\{\check{\tau}_3,\check{g}\}
-\frac{ie}{2m}\vec{A}\cdot\nabla _{R}
\{\check{\tau}_3,\check{g}\}
+ie\vec{v}_{\rm F}\times\vec{B}^{\rm em}\cdot\frac{\partial}{\partial\vec{k}_{\parallel}}\check{g}=0,
\end{align}
\end{widetext}
where $[a,b]=ab-ba$, $\{a,b\}=ab+ba$, $\vec{v}_{\rm F}$ is the Fermi velocity, $\vec{B}=\nabla\times\vec{A}$, $\vec{B}^{\rm em}=\nabla\times \vec{A}^{\rm em}$, and
$\check{\Delta}$ in Eq.(\ref{2-15}) is the Wigner transformation of the gap function, which is given by,
\begin{eqnarray}
\check{\Delta}(\vec{k},\vec{R})=
\left(
\begin{array}{cc}
0 & -\frac{\Delta (\vec{R})}{k_F}(k_x-ik_y) \\
\frac{\Delta^{*} (\vec{R})}{k_F}(k_x+ik_y) & 0
\end{array}
\right).
\end{eqnarray}
Eq.(\ref{2-15}) describes responses of quasiparticle excitations and Cooper pairs to an applied magnetic field and an emergent chiral magnetic field. It is worth noting that in the last term of Eq. (\ref{2-15}), the coupling with the pseudo-Lorentz force due to the chiral magnetic field $\vec{B}^{\rm em}$ appears. 
There is an important difference between this pseudo-Lorentz force term and the usual Lorentz force term, i.e. 
the third term of Eq.(\ref{2-15}) which depends on $\vec{B}$. 
As seen from Eq.(\ref{2-15}), the usual Lorentz force term couples only with the normal Green's functions $g$ and $\bar{g}$ which describe Bogoliubov quasiparticles,
and does not couple to the anomalous Green's functions, $f$ and $f^{\dagger}$ in Eq. (\ref{2-15}), because of $\check{\tau}_3$ in this term.
The physical reason of this feature is understood as follows. Cooper pairs are composed of electrons with momentum $\vec{k}$ and $-\vec{k}$, and hence the Lorentz force $\sim \vec{v}\times\vec{B}\propto \vec{k}\times \vec{B}$ acting on these two electrons cancels with each other.
 In contrast, the pseudo-Lorentz force term due to $\vec{B}^{\rm em}$ couples with both $g$ and $f$ in Eq.(\ref{2-15}),
 and hence, dynamics of Cooper pairs is affected by this term. 
 This is because that two fermions which form a Cooper pair in Weyl superconductors carry chirality with opposite signs as well as
 momentum $\vec{k}$ and $-\vec{k}$, and thus, the pseudo-Lorentz force due to $\vec{B}^{\rm em}$, the sign of which depends on the chirality, does not cancel between
 these two fermions, but instead adds up to twice in magnitude. 
 In the following, we neglect the Lorentz force term due to a real magnetic field, which is irrelevant to Cooper pair dynamics. Furthermore, we consider perturbative expansion up to the first order in terms of the gradient, the vector potential and the chiral magnetic field. Then, we obtain a simplified equation,
\begin{align}
\label{2-16}
[i\omega _n \check{\tau}_3 - \check{\Delta}+e\vec{v}_{\rm F}\cdot\vec{A}\check{\tau}_3, \check{g}]
& +ie\vec{v}_{\rm F}\times\vec{B^{\rm em}}\cdot\frac{\partial}{\partial\vec{k}_{\parallel}}\check{g} \nn \\
& + i\vec{v}_{\rm F}\cdot{\nabla}_{R}\check{g}=0.
\end{align}
Within this approximation, the normalization condition, 
\beq
\check{g}^{2}=\check{1},
\label{eq:norm}
\eeq 
is valid because $\check{g}^2$ also satisfies Eq. (\ref{2-16}). The normalization condition gives $\overline{g}=-g$ and $g^2-ff^\dag=1$. Eq. (\ref{2-16}) and this normalization condition are the basis of the following argument. Here, we comment that the normalization condition for the quasiclassical Green's functions holds, because of the approximation that the gradient terms of the self-energy are negligible.


\subsection{\label{sec:A}Supercurrent flow akin to chiral magnetic effect}
In this subsection, we derive the expression for a supercurrent from the Eilenberger equation (\ref{2-16}),  and demonstrate that supercurrent flow is induced by the emergent chiral magnetic field. We would like to stress that this is not the Meissner current, as clarified below. We expand  the quasiclassical Green's function up to the first order in the spatial gradient, the vector potential and the chiral magnetic field,
\begin{equation}
\label{eq:3-1}
\check{g}=\check{g}_0+\check{g}_1,
\end{equation}
where $\check{g}_0$ is the Green's function for a homogeneous system and $\check{g}_1$ is the first order correction. Solving the zeroth order equation of Eq. (\ref{2-16}) under the normalization condition $g_0^2-f_0f_0^\dag=1,\;\overline{g}_0=-g_0$, we obtain the zeroth order contributions,
\begin{gather}
\label{g0}
g_0=\frac{\omega_n}{\sqrt{\omega_n^2+|\Delta(\vec{R})\sin \theta_k|^2}},\\ \label{f0}
f_0=\frac{\Delta(\vec{R})e^{-i\phi_k}\sin \theta_k}{i\sqrt{\omega_n^2+|\Delta(\vec{R})\sin \theta_k|^2}},\\ \label{f0dagger}
f_0^\dag=\frac{\Delta^*(\vec{R})e^{i\phi_k}\sin \theta_k}{i\sqrt{\omega_n^2+|\Delta(\vec{R})\sin \theta_k|^2}}.
\end{gather}
Here, we assume an isotropic Fermi surface, and introduce polar coordinates in momentum space with $\phi_k$, and $\theta_k$ azimuth and polar angles, respectively. The first order correction terms of the quasiclassical equation (\ref{2-16}) are given by, 
\begin{align}
\label{Eilenburger first1}
-i\vec{v}_{\rm F}\cdot \vec{\partial}f_0-2i\omega_nf_1
&-ie\vec{v}_{\rm F}\times\vec{B}^{\rm em}\frac{\partial}{\partial\vec{k}_{\parallel}}f_0 \nn \\
&+2\Delta (\vec{R})e^{-i\phi_k}\sin \theta_kg_1=0,
\end{align}
and
\begin{align}
\label{Eilenburger first2}
&i\vec{v}_{\rm F}\cdot\vec{\partial}f_0^\dag-2i\omega_nf_1^\dag\nn\\
&+ie\vec{v}_{\rm F}\times\vec{B}^{\rm em}\frac{\partial}{\partial\vec{k_{\parallel}}}f_0^\dag+2\Delta^*(\vec{R}) e^{i\phi_k}\sin \theta_kg_1=0.
\end{align}
Here, we introduce the gauge invariant differential operator,
\begin{eqnarray}
\vec{\partial}\equiv\begin{cases}
\nabla_R-2ie\vec{A}\;\;\;\;\;\;\;{\rm for}\;\;f\;\;{\rm and}\; \Delta, \\
\nabla_R+2ie\vec{A}\;\;\;\;\;\;\;{\rm for}\;\;f^\dag\;{\rm and}\; \Delta^*.
\end{cases}
\end{eqnarray}
Solving Eq. (\ref{Eilenburger first1}) and (\ref{Eilenburger first2}) under the condition $2g_0g_1-f_0f_1^\dag-f_1f_0^\dag=0,\;\overline{g}_1=-g_1$, we obtain
\begin{eqnarray}
g_1=g_1^{\rm M}+g_1^{\rm T},
\end{eqnarray}
\begin{gather}
\label{g1}
g_1^{\rm M} \equiv \frac{i\sin^2 \theta_k}{2(\omega_n^2+|\Delta\sin \theta_k|^2)^{\frac{3}{2}}}{\vec{v}_F\cdot(\nabla_R \chi-2e\vec{A})}|\Delta|^2, \\
\label{g1t}
g_1^{\rm T} \equiv \frac{-ie\sin \theta_k}{2k_F(\omega_n^2+|\Delta\sin \theta_k|^2)^{\frac{3}{2}}}(\vec{v}_F\times \vec{B}^{\rm em})\cdot \vec{\phi}|\Delta|^2,
\end{gather}
where $\chi$ is the phase of the superconducting gap and $\vec{\phi}$ is a unit vector in the $\phi_k$-direction. $g_1^{\rm M}$ is the usual contribution in the superconducting state which gives rise to the Meissner effect. It is noted that this term does not contain the emergent chiral vector potential and thus, the Meissner effects due to the emergent chiral magnetic field does not occur. The second term $g_1^{\rm T}$  which is originated from the emergent chiral magnetic field is a distinct feature of this system. 

The expression of a charge current can be derived from Eq (\ref{g1}) and (\ref{g1t}),
\begin{eqnarray}
\label{sc}
\vec{j}_{\rm s}=-2e\nu(0)\pi iT\sum_{n}\int{}\frac{d\Omega_k}{4\pi}\vec{v}_Fg.
\end{eqnarray}
Here, $\nu(0)$ is the density of states at the  Fermi energy, and $\int d\Omega_k$ is the integral over the direction of the Fermi momentum. 
Since we consider only a static vector potential $\vec{A}$, currents carried by quasiparticles are absent, and there is only a supercurrent. The contribution arising from the pseudo-Lorentz force is given by
\begin{align}
\label{3-2}
\vec{j}_{\rm s}^{\rm T}&=-2e\nu(0)\pi i T\sum_{n}\int{}\frac{d\Omega_k}{4\pi}\vec{v}_Fg_1^{\rm T}\nonumber\\
&=-e^2\nu(0)\pi T\sum_{n}\int{}\frac{d\Omega_k}{4\pi}\vec{v}_{\rm F}
\frac{\sin \theta_k(\vec{v}_{\rm F}\times B^{\rm em})\cdot\vec{\phi}|\Delta|^2}{k_F(\omega_n^2+|\Delta|^2\sin^2 \theta_k)^{\frac{3}{2}}}.
\end{align}
We consider a uniform chiral magnetic field along the $z$-direction. It is noted that the emergent chiral magnetic field $\vec{B}^{\rm em}$ is an odd function of the momentum, i.e. its sign depends on chirality of Weyl points at $\vec{k}=(0,0,\pm k_0)$.
Therefore, it is legitimate to assume that $\vec{B}^{\rm em}=(0,0,B^{\rm em}(\vec{k}))$ satisfies
\begin{eqnarray}
B^{\rm em}(\vec{k})=\begin{cases}
B^{\rm em}\;\;\;\;\;\;\;\;\;\;{\rm for}\;\;\;0\leq\theta_k\leq\Theta\\
-B^{\rm em}\;\;\;\;\;\;\;{\rm for}\;\;\;\pi-\Theta\leq\theta_k \leq \pi\\
0\;\;\;\;\;\;\;\;\;\;\;\;\;\;\;\;\;\; {\rm otherwise},
\end{cases}
\end{eqnarray}
with $B^{\rm em} > 0$ a constant.
Here, $\Theta$ is the cut off of the polar angle, which determines the momentum region where the approximation of the linearized dispersion is valid. Then, Eq (\ref{3-2}) can be written as   
\begin{eqnarray}
\label{torsion current}
\vec{j}_s^{\rm T}&=&\frac{e^2\nu(0)\pi T v_F^2 B^{\rm em}|\Delta|^2}{k_F} \sum_{n}\int_{0}^{\Theta}d\theta_k\frac{\sin^3\theta_k\cos \theta_k}{{(\omega_n^2+|\Delta|^2\sin^2 \theta_k)^{\frac{3}{2}}}}\hat{z},\nonumber\\
\end{eqnarray}
where $\hat{z}$ is unit vector of $z$-direction. 
Eq. (\ref{torsion current}) implies that the chiral magnetic field induces supercurrent flow parallel to the chiral magnetic field. This effect is akin to the CME in Weyl semimetals. However, it is noted that the result shown above is not directly related to chiral anomaly of Weyl quasiparticles, since it arises from the response of Cooper pairs to the chiral magnetic field. 

\begin{figure}[h]
\begin{center}
\includegraphics[width=85mm]{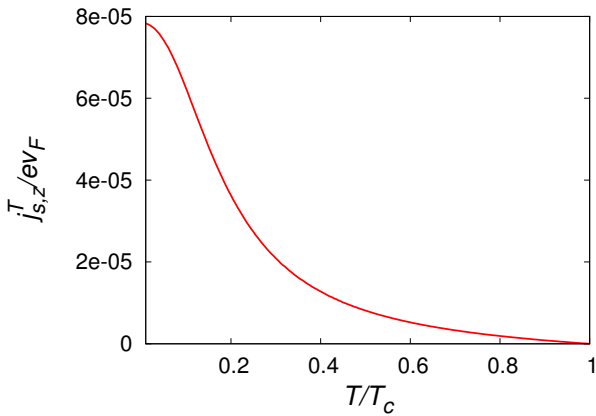}
\end{center}
\vspace{5mm}
\caption{Temperature dependence of a supercurrent  induced by a chiral magnetic field for $eB^{\rm em}=0.2$. 
The cutoff for integral over polar angle is set as $\Theta=\frac{\pi}{10}$. 
The magnitude of the current is divided by $ev_F$.}
\label{fig0}
\end{figure}

To capture qualitative features of the supercurrent flow induced by the chiral magnetic field, we carry out numerical estimation of Eq. (\ref{torsion current}). The calculated result of the supercurrent plotted as a function of temperature is shown in FIG. \ref{fig0}.
In this calculation, we used an approximated temperature dependence of the BCS gap function, $|\Delta(T)|\simeq1.765T_c\tanh \left(1.74\sqrt{\frac{T_c}{T}-1}\right)$ with $T_c$ a critical temperature.\cite{Tinkham} 
The physical origin of this effect is understood as follows. Two electrons which constitute a Cooper pair in WSCs carry opposite chiralities as well as opposite signs of momenta. Thus, the pseudo-Lorentz force due to the chiral magnetic field which depends on chirality does not cancel between these two electrons, in contrast to the usual Lorentz force, which does not depend on chirality.
As a result, the pseudo-Lorentz force gives rise to a supercurrent.
We stress again that this supercurrent is not a Meissner current, because the Meissner effect for the chiral magnetic effect is absent, as 
shown in Eq.(\ref{g1}). 
Although the analysis shown above is based on a single-band toy model in which the chiral magnetic field is introduced by hand, we will verify, in the Sec.\ref{sec:level4}, that in a realistic superconducting Dirac semimetal model, supercurrent akin to the CME can actually be generated by lattice torsion.

\subsection{\label{sec:B}Ginzburg-Landau equation and the Fulde-Ferrell state}
In the previous subsection,
it is found that the pseudo-Lorentz force induces a supercurrent which flows parallel to the chiral magnetic field. This implies that when the bulk supercurrent flow is prohibited by a boundary condition of the system, spatial modulation of the superconducting phase occurs and the Fulde-Ferrell state is realized. To confirm this prediction, we, here, derive the GL equation microscopically from the Eilenberger equation. To derive the GL equation, we expand the quasiclassical Green's functions $g$ and $f$ up to the third order in $|\Delta|$ and the second order in the spatial gradient. \cite{Kopnin} Applying the gradient expansion up to the second order,
\begin{eqnarray}
g=g^{(0)}+g^{(1)} +g^{(2)},
\end{eqnarray}
\begin{eqnarray}
\label{f012}
f=f^{(0)}+f^{(1)} +f^{(2)},
\end{eqnarray}
where $g^{(i)}$ and $f^{(i)}$ are the $i$-th order corrections with respect to the spatial gradient, we rewrite the Eilenberger equations up to the first order in the following form,
\begin{align}
\label{4-2-1}
-i\vec{v}_{\rm F}\cdot\vec{\partial}f^{(0)}&-2i\omega_n f^{(1)}-ie\vec{v}_F\times\vec{B}^{\rm em}
\cdot\frac{\partial}{\partial \vec{k}_{\parallel}}f^{(0)}\nonumber\\
&+2\Delta e^{-i\phi_k}\sin \theta_k g^{(1)}=0,
\end{align}
\begin{align}
\label{4-2-2}
i\vec{v}_{\rm F}\cdot\vec{\partial}f^{\dag (0)}
&-2i\omega_n f^{\dag(1)}+ie\vec{v}_F\times\vec{B}^{\rm em}\cdot\frac{\partial}{\partial \vec{k}_{\parallel}}f^{\dag (0)}\nonumber\\
&+2\Delta^* e^{i\phi_k}\sin \theta_k g^{(1)}=0.
\end{align}
Note that $\vec{B}^{\rm em}$ is the first order in the spatial derivative. Solving Eqs. (\ref{4-2-1}) and (\ref{4-2-2}) under the normalization condition $2g^{(0)}g^{(1)}=f^{(0)}f^{\dag(1)}+f^{(1)} f^{\dag (0)}$, we obtain the first order corrections, 
\begin{gather}
\label{4-3}
g^{(1)}=o(|\Delta|^2), \\
\label{f1}
f^{(1)}=\frac{-\vec{v}_F\cdot\vec{\partial}f^{(0)}-e\vec{v}_{\rm F}
\times\vec{B}^{\rm em}\cdot\frac{\partial}{\partial\vec{k}_{\parallel}}f^{(0)}}{2\omega_n}.
\end{gather}
In a similar manner, we can obtain the Eilenberger equations for the second order corrections, and their solutions given by,
\begin{gather}
\label{eq:4-6}
g^{(2)}=o(|\Delta|^2), \\
\label{f2}
f^{(2)}=\frac{-\vec{v}_{\rm F}\cdot\vec{\partial}f^{(1)}-e\vec{v}_{\rm F}
\times\vec{B}^{\rm em}\cdot\frac{\partial}{\partial\vec{k}_{\parallel}}f^{(1)}}{2\omega_n}.
\end{gather}
According to Eqs.~(\ref{g0})-(\ref{f0dagger}), we expand $g^{(0)}$ and $f^{(0)}$ with respect to the gap function $|\Delta|$ up to the third order,
\begin{gather}
\label{4-1}
g^{(0)}=1-\frac{|\Delta(\vec{x})|^2\sin^2 \theta_k}{2\omega_n^2}, \\
\label{4-1-2}
f^{(0)}=\frac{\Delta(\vec{x})e^{-i\phi_k}\sin \theta_k}{i|\omega_n|}-\frac{|\Delta(\vec{x})|^2\Delta(\vec{x}) e^{-i\phi_k}\sin^3 \theta_k }{2i|\omega_n|^3}, \\
\label{4-1-3}
f^{\dag (0)}=\frac{\Delta(\vec{x})^*e^{i\phi_k}\sin \theta_k}{i|\omega_n|}-\frac{|\Delta(\vec{x})|^2\Delta(\vec{x})^* e^{i\phi_k}\sin^3 \theta_k}{2i|\omega_n|^3}. 
\end{gather}
Substituting Eqs. (\ref{4-1})-(\ref{4-1-3}) into Eqs. (\ref{f1}) and (\ref{f2}), we obtain correction terms of the anomalous Green's function which are necessary for the derivation of the GL equation.

We assume an effective pairing interaction for a $p+ip$ wave superconductor in the following form,
\begin{eqnarray}
V(\hat{k},\hat{k}')=-|V_0|(\hat{k}_x-i\hat{k}_y)(\hat{k}'_x+i\hat{k}'_y).
\end{eqnarray}
$|V_0|$ is the strength of the effective attraction. Then, the gap equation can be written as
\begin{align}
\frac{\Delta(\vec{k}_{\parallel},\vec{x})}{\lambda}
&=\pi iT \sum_{n}\int{}\frac{d\Omega_{k'}}{4\pi} 
(\hat{k}_x-i\hat{k}_y)(\hat{k}'_x+i\hat{k}'_y)f(\vec{k}_{\parallel}',\vec{x})\nonumber\\
&=\langle e^{i\phi_{k}}\sin \theta_kf(\vec{k}_{\parallel},\vec{x}) \rangle_{\rm F} e^{-i\phi_k}\sin \theta_k
\end{align}
where $\lambda=|V_0|\nu(0)$ is the dimensionless coupling constant and $\langle\cdots\rangle_{\rm F}$ means the average over the Fermi surface.  From Eqs.~(\ref{f012}) and (\ref{4-1-2}), we have, 
\begin{widetext}
\begin{align}
\label{eq:4-7}
\langle e^{i\phi_{k}}\sin \theta_kf(\vec{k}_{\parallel},\vec{x}) \rangle_{\rm F}
&=\left\langle \frac{\Delta(\vec{x})\sin \theta_k}{i|\omega_n|}
-\frac{|\Delta(\vec{x})|^2\Delta(\vec{x})\sin^3 \theta_k }{2i|\omega_n|^3}+e^{i\phi_{k}}
\sin \theta_kf^{(1)}+e^{i\phi_{k}}\sin \theta_kf^{(2)} \right\rangle_{\rm F} \nn \\
&=\frac{2\Delta(\vec{x})}{3i|\omega_n|}-\frac{4|\Delta(\vec{x})|^2\Delta(\vec{x})}{15i|\omega_n|^3}+\langle e^{i\phi_{k}}\sin \theta_kf^{(1)} \rangle_{\rm F}+\langle e^{i\phi_{k}}\sin \theta_kf^{(2)} \rangle_{\rm F}.
\end{align}
Using Eqs. (\ref{f1}), (\ref{f2}) and (\ref{4-1-2}), we perform the average over the Fermi surface and the sum of the Matsubara frequency. Then, up to the  third order with respect to $|\Delta|$ and the second order in the spatial gradient, we obtain the GL equation, 
\begin{align}
\label{GL}
\left(1-\frac{T}{T_c}\right)\Delta(\vec{x})-\frac{7\zeta(3)}{10\pi^2 T_{c}^2}|\Delta(\vec{x})|^2 \Delta(\vec{x})+\frac{7\zeta(3)v_{\rm F}^2}{40\pi^2 T_{c}^2}\left(\partial_x^2+\partial_y^2+\frac{1}{2}\partial_z^2\right)\Delta(\vec{x})+\frac{7\zeta(3)v_F^2}{64\pi^2 T_c^2}\left(3i\frac{eB^{em}}{k_F}\partial_z-4\left( \frac{eB^{em}}{k_F} \right)^2\right)\Delta(\vec{x})=0.
\end{align}
\end{widetext}
The first order gradient term which is linear in $B^{\rm em}$ implies that the Fulde-Ferrell state with nonzero center of mass momentum of Cooper pairs is stabilized by the emergent chiral magnetic field. For the gap function, 
\beq
\Delta(\vec{x})=\Delta e^{iQz}, 
\eeq 
we determine the value of $Q$ by maximizing the critical temperature $T_c$, and obtain $Q=-\frac{15eB^{\rm em}}{8k_F}$. 
It is found that $T_c$ is decreased as $B^{\rm em}$ increases, because of the $(B^{\rm em})^2$ term in the GL equation, though the $B^{\rm em}$-linear term increases $T_c$. 

The realization of the Fulde-Ferrell state is in accordance with the result in the previous subsection for a supercurrent induced by the chiral magnetic field, since the spatially modulated phase implies supercurrent flow along the direction of the modulation provided that
the current flow is not forbidden by a boundary condition.

\section{\label{sec:level4}Dirac superconductivity realized in Dirac metal model with lattice torsion}
In this section, we consider a DSM model with lattice torsion, and the Dirac superconducting state realized in this system. We exploit a two-orbital model for  real DSM materials such as Cd$_3$As$_2$ and Ni$_3$Pb. In this model, because of orbital-texture on the Fermi surface, spin-triplet and orbital-singlet pairing states are
stabilized.~\cite{TSC Dirac,SC Dirac} Using the quasiclassical Eilenberger equation for this model, we find that
although the lattice torsion does not induce charge supercurrents, because of time-reversal symmetry,
it generates spin supercurrent flow parallel to an emergent chiral magnetic field.

\subsection{\label{sec A} C$_4$ symmetric Dirac semimetal model}
\begin{figure}[h]
\begin{center}
\includegraphics[width=82mm]{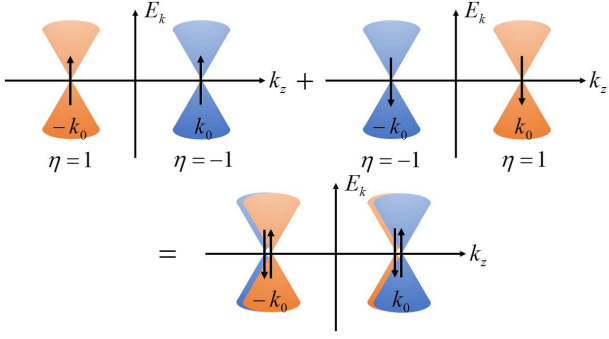}
\end{center}
\caption{Spin-up and spin-down Dirac fermions in topological DSMs. Here, $\eta$ denotes chirality of each Weyl cone, and the arrows denote the spin states. }
\label{sDirac}
\end{figure}

DSMs have Dirac points in the bulk momentum space and its low energy physics is described by the Dirac Hamiltonian. DSMs have both time-reversal and inversion symmetries. The minimal effective Hamiltonian of these systems is constructed from at least four degrees of freedom, e.g. two spin and two orbital degrees of freedom. The Kramers theorem guarantees doubly degenerate Bloch states. As mentioned before,  the stability of Dirac points requires symmetry protection mechanism. In the case of Cd$_3$As$_2$ and Ni$_3$Pb, $C_4$ rotational symmetry protects the Dirac points,  which is characterized by $C_4$ topological invariant.\cite{Fang} 
In the basis set $\{\ket{P_\frac{3}{2},\frac{3}{2}}, \ket{S_\frac{1}{2},\frac{1}{2}}, \ket{P_\frac{3}{2},-\frac{3}{2}},\ket{S_\frac{1}{2},-\frac{1}{2}} \}$, 
the effective Hamiltonian for this topological DSMs near the $\Gamma$ point \cite{franz, Liu} can be written as,
\begin{align}
\label{5-1}
H^N(\vec{k})&=\begin{pmatrix} 
h^{\rm latt}(\vec{k})&&0\\
0&&(h^{\rm latt }(-\vec{k}))^{*}
\end{pmatrix}
\end{align}
where 
$h^{\rm latt}(\vec{k})=m_k\sigma_z+\Lambda(\sigma_x \sin ak_x+\sigma_y \sin ak_y)$, $m_k=t_0+t_1 \cos ak_z +t_2 (\cos ak_x+\cos ak_y)$, and $\sigma_j\;(j=1,2,3)$
is the Pauli matrix for orbital space.
The effective Hamiltonian is diagonal in spin space,  and in the energy spectrum, there are Dirac points at $\vec{k}^{\eta}=(0,0,\eta \arccos(-\frac{t_0+2t_2}{t_1}))\equiv(0,0,\eta k_0)$, as shown in FIG.~\ref{sDirac}. 
Here, $\eta=\pm 1$ is chirality of Dirac fermions, i.e. a sign of a monopole charge in the momentum space.
The expansion of the upper-left block of the effective Hamiltonian $h^{\rm latt}(\vec{k})$ in the vicinity of the Dirac points $\vec{k}^\eta$ up to the second order in the momentum can be written as,
\begin{eqnarray}
h(\vec{k}-\vec{k}^\eta)&&=v_j^\eta\sigma_j(k_j-k^\eta_j)\nonumber\\&&-\frac{1}{2M_{ij}}(k_i-k^\eta_i)(k_j-k^\eta_j)\sigma_z,
\end{eqnarray}
where $\vec{v}^{\eta}\equiv(\Lambda a,\Lambda a, -\eta t_1 a \sin ak_0)$ is the Fermi velocity and $M_{ij}$ is the effective mass tensor which is given by
\begin{eqnarray}
(M_{ij})^{-1}\equiv 
\begin{cases}
0  & \mbox{for  $i\neq j$} \\
a^2t_2 & \mbox{for $i=j=x$, $y$} \\
a^2t_1\cos a k_0  & \mbox{for $i=j=z$} 
\end{cases}.
\end{eqnarray}

The leading term of the mechanical strain can be incorporated in the lattice model by modifying the hopping amplitude along the $z$-direction,\cite{franz}
\begin{eqnarray}
\label{0-2}
t_1\sigma_z\rightarrow t_1(1-u_{33})\sigma_z+i\Lambda(u_{13}\sigma_x+u_{23}\sigma_y).
\end{eqnarray}
Here, $u_{ij}=(\partial_i u_j+\partial_j u_i)/2$ is the symmetric strain tensor, and $\vec{u} = (u_1, u_2, u_3)$ represents the displacement vector. Within the leading order in terms of momentum, the mechanical strain behaves as a chiral vector potential. The upper-left block of the effective Hamiltonian for a distorted crystal is given by,
\begin{align}
h(\vec{k}-&\vec{k}^\eta)=v_j^\eta\sigma_j(k_j-k^\eta_j-eA_j^{\rm em}(\eta))\nonumber\\
&-\frac{1}{2M_{ij}}(k_i-k^\eta_i)(k_j-k^\eta_j)\sigma_z=\Gamma^{\eta}_j(\vec{k})\sigma_j.
\end{align}
The Gor'kov equations for the superconducting state of the DSM model have the form similar to Eqs.(\ref{eq:gor1}) and (\ref{eq:gor2}),
but with matrices defined in the $8\times 8$ spin, orbital, and particle-hole spaces.
In this case, the Green's function $\check{G}$ is  defined by Eq.(\ref{eq:green1}) with,
\begin{eqnarray}
G(x_1,x_2)=
\left(\begin{array}{cc}
G_{\uparrow} & 0 \\
0& G_{\downarrow}
\end{array}
\right), ~
F(x_1,x_2)=
\left(\begin{array}{cc}
F_{\uparrow} & 0 \\
0& F_{\downarrow}
\end{array}
\right),  ~\mbox{etc.} \nonumber
\end{eqnarray}
Here,
\begin{eqnarray}
G_{\uparrow}(x_1,x_2)=
\left(\begin{array}{cc}
G_{P\frac{3}{2},P\frac{3}{2}} & G_{P\frac{3}{2},S\frac{1}{2}} \\
G_{S\frac{1}{2},P\frac{3}{2}}  & G_{S\frac{1}{2},S\frac{1}{2}}
\end{array}
\right),
\end{eqnarray}
\begin{eqnarray}
G_{\downarrow}(x_1,x_2)=
\left(\begin{array}{cc}
G_{P-\frac{3}{2},P-\frac{3}{2}} & G_{P-\frac{3}{2},S-\frac{1}{2}} \\
G_{S-\frac{1}{2},P-\frac{3}{2}}  & G_{S-\frac{1}{2},S-\frac{1}{2}}
\end{array}
\right),
\end{eqnarray}
with
\begin{eqnarray}
G_{\alpha\beta}(x_1,x_2)=\langle T_\tau\psi_{\alpha}(x_1)\psi_{\beta}^\dag(x_2) \rangle ,
\end{eqnarray}
and $\alpha,\beta = \{ P\frac{3}{2}, S\frac{1}{2}, P-\frac{3}{2}, S-\frac{1}{2} \}$.
The other Green's functions, $F_{\uparrow}$, $F_{\downarrow}$, $\bar{G}_{\uparrow}$ and  $\bar{G}_{\downarrow}$, are defined in a similar way.
The kinetic energy term $\check{\zeta}$ in the Gor'kov equation is given by,
\begin{widetext}
\begin{eqnarray}
\label{9-2}
\check{\zeta}=\begin{pmatrix}
\Gamma_j^\eta(\vec{k})\sigma_j-\mu&&0&&0&&0\\
0&&\Gamma_j^{-\eta}(-\vec{k})\sigma_j^*-\mu&&0&&0\\
0&&0&&\Gamma_j^{-\eta}(-\vec{k})\sigma_j^*-\mu&&0\\
0&&0&&0&&\Gamma_j^\eta(\vec{k})\sigma_j-\mu
\end{pmatrix},
\end{eqnarray}
\end{widetext}
where $\Gamma^\eta_j(\vec{k})\equiv v_j^\eta q_j-\frac{1}{2M_{lm}}(k_l-k^\eta_l)(k_m-k^\eta_m)\delta_{jz}$ and $q_j=k_j-k^\eta_j-eA_j^{\rm em}(\eta)$. The chiral vector potential generated by mechanical strain reads,
\begin{eqnarray}
\label{0-4}
\vec{A}^{\rm em}(\eta)=-\frac{\eta}{ea}(u_{13}\sin ak_0,u_{23}\sin ak_0,u_{33} \cot ak_0).
\end{eqnarray}
From now on,  for simplicity, we assume a uniform chiral magnetic field, which can be
realized by twisting topological DSMs.~\cite{franz} In this case, the chiral vector potential and the chiral magnetic field can be, respectively, written as $\vec{A}^{\rm em}(\eta)=\eta B^{\rm em}(-y,x,0),\; \vec{B}^{\rm em}(\vec{k^\eta})=(0,0,\eta B^{\rm em})$. 
\begin{figure}[h]
\begin{center}
\includegraphics[width=30mm]{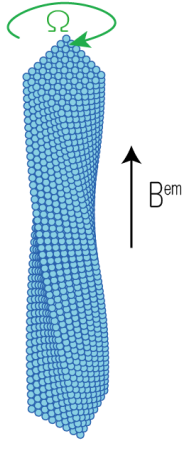}
\end{center}
\caption{Emergent chiral magnetic field generated by torsion.}
\label{pseudo magnetic}
\end{figure}

According to Refs.~\onlinecite{TSC Dirac} and \onlinecite{SC Dirac}, in this DSM model, a spin-triplet, orbital-singlet pairing state rather than the ordinary $s$-wave pairing state is favored because of its momentum-orbital texture on the Fermi surface. The order-parameter matrix of this pairing state
for the same basis as Eq.(\ref{9-2})
 is given by, 
\begin{eqnarray}
\label{9-1}
\check{\Delta}=\begin{pmatrix}
0&&0&&-i\sigma_y \Delta&&0\\
0&&0&&0&&-i\sigma_y\Delta\\
-i\sigma_y \Delta^*&&0&&0&&0\\
0&&-i\sigma_y \Delta^*&&0&&0
\end{pmatrix}.
\end{eqnarray}
This superconducting gap also preserves $C_4$ symmetry. This guarantees the existence of the point nodes in the energy spectrum of the Bogoliubov quasiparticles, which characterize the Dirac superconducting state. Since both of the kinetic energy matrix (\ref{9-2}) and the order-parameter matrix (\ref{9-1}) are block-diagonalized in spin space, 
we can deal with each spin sector separately. 
In the following, we denote the spin-up (spin-down) states corresponding to $\{|P,\frac{3}{2}\rangle,|S,\frac{1}{2}\rangle \}$
($\{|P,-\frac{3}{2}\rangle,|S,-\frac{1}{2}\rangle \}$)
as the $s_z=1$ ($-1$) state. 
The Nambu Hamiltonian of each spin sector can be written as, 
\begin{eqnarray}
\label{9-3}
\check{H}^{s_z=1\eta}=\begin{pmatrix}
\Gamma_j^\eta(\vec{k})\sigma_j-\mu&&-i\sigma_y\Delta\\
-i\sigma_y\Delta^*&&\Gamma_j^{-\eta}(-\vec{k})\sigma_j^*-\mu
\end{pmatrix},
\end{eqnarray}
\begin{eqnarray}
\label{9-4}
\check{H}^{s_z=-1\eta}=\begin{pmatrix}
\Gamma_j^{-\eta}(-\vec{k})\sigma_j^*-\mu&&-i\sigma_y\Delta\\
-i\sigma_y\Delta^*&&\Gamma_j^\eta(\vec{k})\sigma_j-\mu
\end{pmatrix}.
\end{eqnarray} 
To proceed further, we introduce parameters $\theta^\eta$ and $\phi^\eta$ defined as,
\begin{align}
\cos \theta^\eta &\equiv \frac{\Gamma_z^\eta}{\Gamma^\eta},\\
\sin \theta^\eta \cos \phi^\eta &\equiv \frac{\Gamma_x^\eta}{\Gamma^\eta},\\
\sin \theta^\eta \sin \phi^\eta &\equiv \frac{\Gamma_y^\eta}{\Gamma^\eta},
\end{align}
where $\Gamma^\eta \equiv |\vec{\Gamma}^\eta |$
is the energy of the quasiparticle in the normal state. Performing the unitary transformation from the orbital basis to the band basis which diagonalizes the normal state Hamiltonian, we obtain the total Hamiltonian represented by the band basis,
\begin{align}
\label{9-5}
&\check{H}^{s_z=1\eta}\nonumber\\
&\rightarrow \begin{pmatrix}
{\rm diag}(\Gamma^\eta-\mu,-\Gamma^\eta-\mu)&&\Delta \Lambda^\eta\\
-\Delta^* \Lambda^{\eta*}&&{\rm diag}(|\Gamma^\eta|-\mu,-|\Gamma^\eta|-\mu)
\end{pmatrix},\nonumber
\end{align}
and
\begin{align}
&\check{H}^{s_z=-1\eta}\nonumber\\
&\rightarrow \begin{pmatrix}
{\rm diag}(\Gamma^\eta-\mu,-\Gamma^\eta-\mu)&&-\Delta \Lambda^{\eta*}\\
\Delta^* \Lambda^{\eta}&&{\rm diag}(\Gamma^\eta-\mu,-\Gamma^\eta-\mu)
\end{pmatrix}. \nonumber
\end{align}
Here, the matrix $\Lambda^\eta$ is defined by,
\begin{eqnarray}
\Lambda^\eta=\begin{pmatrix}
e^{i\phi^\eta}\sin \theta^\eta&&-e^{i\phi^\eta}\cos \theta^\eta\\
-e^{i\phi^\eta}\cos \theta^\eta&&-e^{i\phi^\eta}\sin \theta^\eta
\end{pmatrix}.
\end{eqnarray}
We assume $\mu>0$, and neglect the lower bands which do not cross the Fermi level. The simplified Nambu Hamiltonian of each spin sector can be written as,
\begin{eqnarray}
\label{eq:dsmh}
\check{H}^{s_z\eta}_{2\times 2}=\begin{pmatrix}
\Gamma^\eta-\mu&&s_z\Delta e^{is_z\phi^\eta}\sin \theta^\eta\\
-s_z\Delta^* e^{-is_z\phi^\eta} \sin \theta^\eta&&\Gamma^\eta-\mu
\end{pmatrix}.\nonumber\\
\end{eqnarray}
This expression is similar to that of the chiral $p$-wave model. However, the spin-up $p+ip$ state and the spin-down $p-ip$ state constitute the Kramers pair, as in the case of the helical $p$-wave or planar state, because of time-reversal symmetry.
\subsection{\label{sec B} Quasiclassical equation for DSM model}
We apply the quasiclassical approximation to this model and derive the Eilenberger equation for the DSM model from the Hamiltonian (\ref{eq:dsmh}). The procedure is similar to the case of Weyl superconductors in Sec.{\ref{sec:level2}}. The quasiclassical Green's function $\check{g}^{s_z\eta}$ for the band with spin $s_z$ and the Fermi surface surrounding the Dirac point at $\vec{k}^{\eta}$ satisfies the Eilenberger equation,
\begin{widetext}
\beq
\left[
i\omega_n\check{\tau}_3-\check{\Delta}^{s_z\eta}(\vec{k},\vec{R}), \check{g}^{s_z\eta}
\right]
+i\vec{V}_{\rm F}^\eta\cdot {\nabla}_R \check{g}^{s_z\eta}+ie\vec{V}_{\rm F}^\eta\times \vec{B}^{\rm em}\cdot\frac{\partial \check{g}^{s_z\eta}}{\partial \vec{k}_{\parallel}}=0,
\label{eq:9-6}
\eeq
where $\vec{V}_F^{\eta}=\vec{V}^{\eta}(\vec{k})|_{\vec{k}=\vec{k}_F}$ is the Fermi velocity which is given by,
\beq
\label{9-7}
\vec{V}^\eta(\vec{k}) &&\equiv \frac{\partial \Gamma^\eta}{\partial \vec{k}}=\frac{1}{\Gamma^\eta}
\left[v_x^{\eta \; 2}q_x-\left\{ v_z^\eta q_z-\frac{1}{2M_{ij}}(k_i-k_i^\eta)(k_j-k_j^\eta)\right\} \frac{k_x}{M_{xx}}, \right. \nn\\
&& \left. v_y^{\eta \; 2}q_y-\left\{ v_z^\eta q_z-\frac{1}{2M_{ij}}(k_i-k_i^\eta)(k_j-k_j^\eta)\right\} 
\frac{k_y}{M_{xx}},\left( v_z^\eta q_z-\frac{1}{2M_{ij}}(k_i-k_i^\eta)(k_j-k_j^\eta) \right)\left(v_z^\eta -\frac{1}{M_{zz}}(k_z-k^\eta_z)\right)\right].\nn \\
\eeq
\end{widetext}
Eq.~(\ref{eq:9-6}) implies that the pseudo-Lorentz force due to a chiral magnetic field exists in the DSM model, and affects Cooper pair dynamics. Here, we note that the normalization condition for the quasiclassical Green's functions holds, because we use the approximation that the gradient terms of the magnetic field and the self-energy are negligible. 

\begin{figure*}[]
\includegraphics[width=65mm]{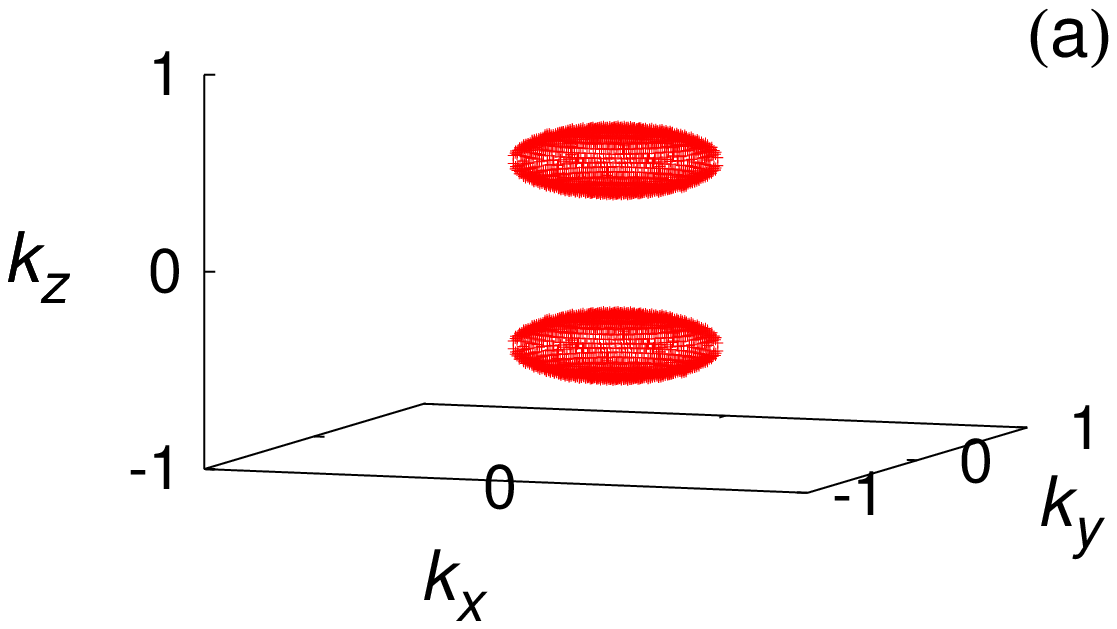}
\vspace{-5mm}
\includegraphics[width=65mm]{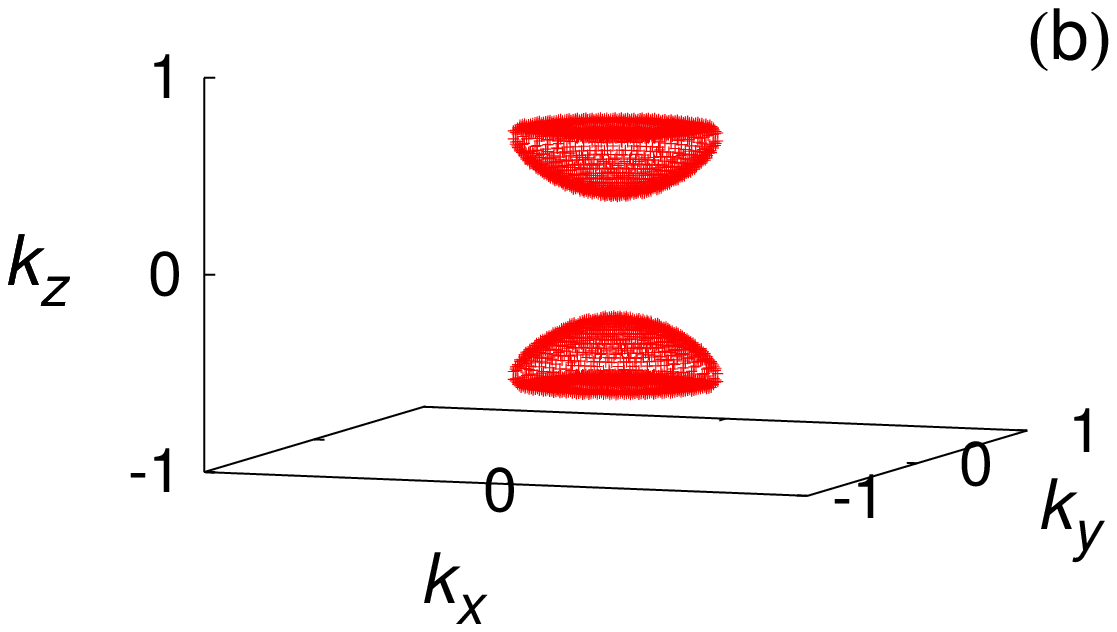}
\includegraphics[width=65mm]{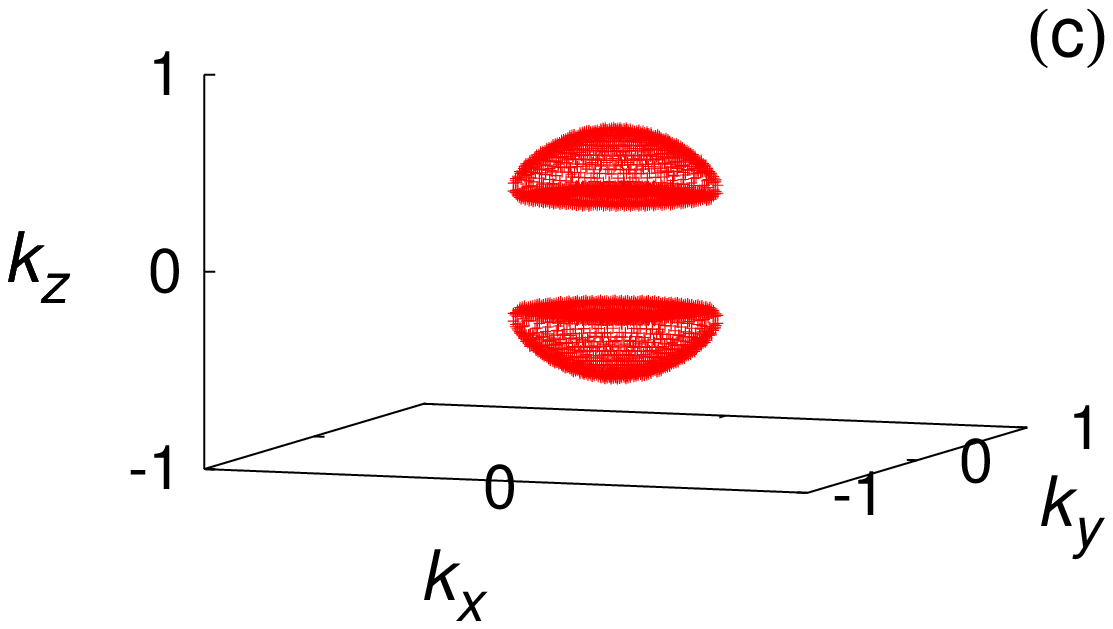}
\includegraphics[width=65mm]{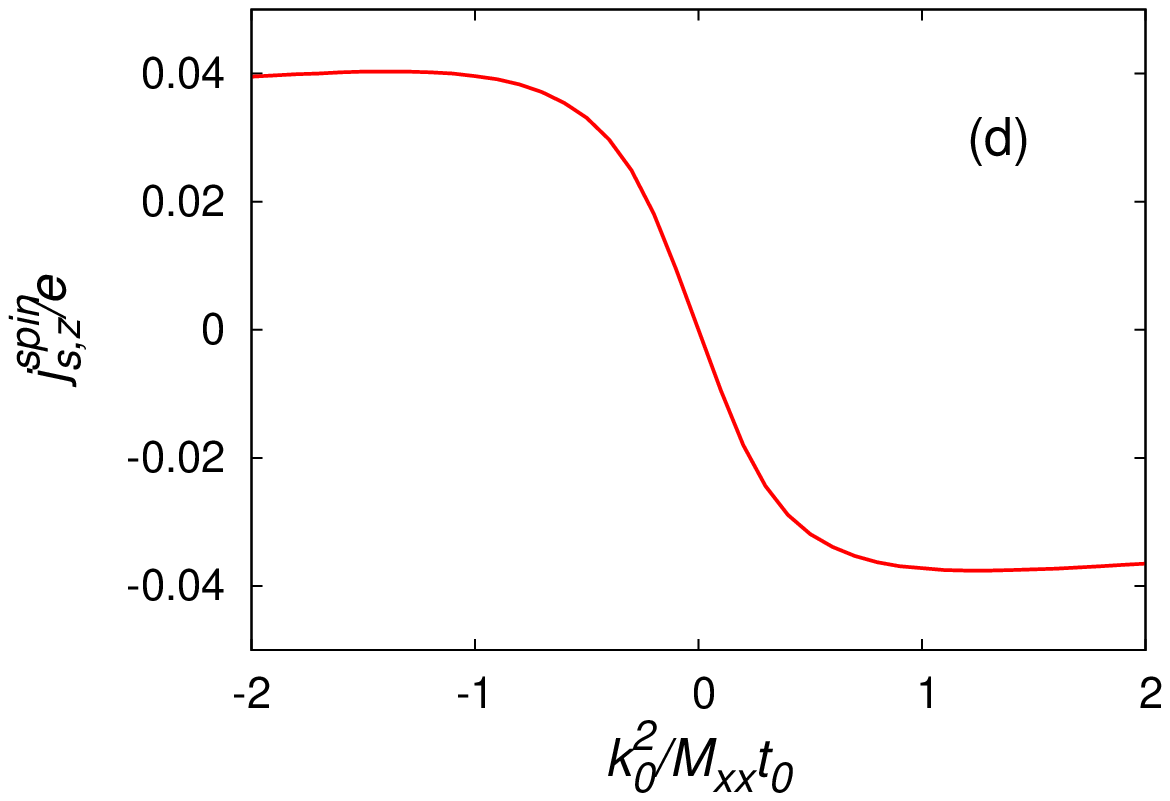}
\centering
\vspace{7mm}
\caption{($a$)-($c$) Fermi surfaces of the DSM model for different values of the effective mass: ($a$) $\frac{k_0^2}{M_{xx}t_0}=0$, ($b$) $\frac{k_0^2}{M_{xx}t_0}=1.0$, and ($c$) $\frac{k_0^2}{M_{xx}t_0}=-1.0$. The other parameters are set to fit the energy dispersion of Cd$_3$As$_2$ as follows: $t_0=-0.2$ eV, $t_1=0.1$ eV, $t_2= 0.07$ eV, $\Lambda=0.040$ eV. Also, we put  $a=4.0$ \AA, $\mu=0.05$ eV. ($d$) Spin supercurrent plotted as a function of anisotropy of the effective mass  $1/M_{xx}$. We put  $B^{em}=0.2$ T, temperature $T=0.756 T_c$ with a critical temperature $T_c=0.0002$ eV,
and the superconducting gap $\Delta=0.000266 $ eV. The other parameters are the same as those used in ($a$)-($c$).
In this plot, the magnitude of the spin supercurrent is divided by the elementary charge $e$.}\label{ani}
\end{figure*}

\subsection{\label{sec C} Spin supercurrent akin to the CME}

The Eilenberger equation for each spin sector obtained in the previous subsection implies that supercurrents can be induced by the torsion-induced chiral magnetic field. However, because of time reversal symmetry, supercurrents carried by spin-up Cooper pairs and spin-down Cooper pairs propagate in the opposite directions. Thus, the net supercurrent does not carry charge but spin; i.e. the nonzero spin supercurrent flows along the chiral magnetic field.

To confirm this prediction, we calculate a spin supercurrent using the Eilenberger equation (\ref{eq:9-6}) in a way similar to that in the previous section. We expand the quasiclassical Green's function up to the first order in the spatial gradient and a chiral magnetic field.
\begin{eqnarray}
\check{g}^{s_z\eta}=\check{g}_0^{s_z\eta}+\check{g}_1^{s_z\eta}
\end{eqnarray}
Solving the zeroth order equation of Eq.~(\ref{eq:9-6}) under the normalization condition $\overline{g}_0^{s_z\eta}\!=\!-g^{s_z\eta}_0$, $g^{s_z\eta\;2}_0-f^{s_z\eta}_0f^{s_z\eta\;\dag}_0\!=\!1$, we obtain the zeroth order terms of the Green's functions,
\begin{eqnarray}
\label{9-10}
g_0^{s_z\eta}=\frac{\omega_n}{\sqrt{\omega_n^2+|\Delta|^2\sin^2 \theta^\eta}},\\
f_0^{s_z\eta}=\frac{is_z\Delta e^{is_z\phi^\eta}\sin \theta^\eta}{ \sqrt{\omega_n^2+|\Delta|^2\sin^2 \theta^\eta}},\\
f_0^{s_z\eta \; \dag}=\frac{is_z\Delta^* e^{-is_z\phi^\eta}\sin \theta^\eta}{ \sqrt{\omega_n^2+|\Delta|^2\sin^2 \theta^\eta}}.
\end{eqnarray}
The first order corrections for the Green's functions satisfy, 
\begin{eqnarray}
\label{9-10-1}
&&2i\omega_n f_1^{s_z\eta}+s_z\Delta e^{i s_z\phi^\eta}\sin \theta^\eta(g_1^{s_z\eta}-\overline{g}_1^{s_z\eta})\nonumber\\&&\;\;+i\vec{V}_F^\eta\cdot\nabla_R f_0^{s_z\eta}+ie\vec{V}_F^\eta \times \vec{B}^{\rm em}\cdot\nabla_{k_\parallel}f_0^{s_z\eta}=0,
\end{eqnarray}
and
\begin{eqnarray}
\label{9-10-2}
&&2i\omega_n f_1^{s_z\eta \; \dag}+s_z\Delta^* e^{-is_z\phi^\eta}\sin \theta^\eta(g_1^{s_z\eta}-\overline{g}_1^{s_z\eta})\nonumber\\&&\;\;-i\vec{V}_F^\eta\cdot\nabla_R  (f_0^{s_z\eta})^{\dagger}-ie\vec{V}_F^\eta \times \vec{B}^{\rm em}\cdot \nabla_{k_\parallel}  (f_0^{s_z\eta})^{\dagger}=0.
\end{eqnarray}
Solving Eq. (\ref{9-10-1}) and Eq. (\ref{9-10-2}) under the normalization condition $2g_0^{s_z\eta}g_1^{s_z\eta}-f_0^{s\eta}f_1^{s_z\eta\; \dag}-f_1^{s_z\eta}f_0^{s_z\eta\; \dag}=0, \; \overline{g}_1^{s_z\eta}=-g_1^{s_z\eta}$, we obtain the first order correction term $g_1^{s_z\eta}$ as
\begin{eqnarray}
\label{9-11}
g_1^{s_z\eta}=g_1^{s_z\eta\;M}+g_1^{s_z\eta\;T},
\end{eqnarray}
\begin{eqnarray}
\label{9-12-1}
g_1^{s_z\eta \;M}\equiv \frac{i}{2}\frac{\vec{V}_F^\eta\cdot\nabla_R \chi}{(\omega_n^2+|\Delta|^2\sin^2 \theta^\eta)^{3/2}}|\Delta|^2,
\end{eqnarray}
\begin{eqnarray}
\label{9-12-2}
g_1^{s_z\eta \; T}\equiv \frac{i}{2}\frac{s_z\vec{V}_F^\eta\cdot\frac{\partial \phi^\eta}{\partial \vec{R}}+es_z\vec{V}_F^\eta \times \vec{B}^{\rm em}\cdot\frac{\partial \phi^\eta}{\partial \vec{k}_{\parallel}}}{(\omega_n^2+|\Delta|^2\sin^2 \theta^\eta)^{3/2}}|\Delta|^2,
\end{eqnarray}
where $\chi$ is the phase of the superconducting gap, and, 
\begin{eqnarray}
\frac{\partial \phi^\eta}{\partial \vec{R}}
=\left(\frac{e\eta B^{\rm em}k_x}{k_x^2+k_y^2},\frac{e\eta B^{\rm em}k_y}{k_x^2+k_y^2},0\right).
\end{eqnarray}

Eq. (\ref{9-12-1}) is a conventional contribution from the phase gradient which generates usual supercurrent flow. The second term Eq. (\ref{9-12-2}) is originated from the pseudo-Lorentz force due to a torsion-induced chiral magnetic field. Since this term is proportional to $s_z$, it does not contribute to total charge currents. However, it can induce a spin supercurrent which is defined by,
\begin{eqnarray}
\vec{j}_{\rm s}^{\rm spin}\equiv \frac{j_{\rm s}^\uparrow-j_{\rm s}^\downarrow}{2(-e)}.
\end{eqnarray}
Then, we obtain,
\begin{eqnarray}
\label{ssc}
\vec{j}_{\rm s}^{\rm spin}&=&\nu(0)\pi iT\sum_{n}\sum_{s_z\eta} \int
\frac{d\Omega_k}{4\pi}s_z\vec{V}_{\rm F}^\eta g_1^{s_z\eta}\nonumber\\
&=&-\nu(0)\pi T\sum_{n}\sum_{\eta} \int_{} \frac{d\Omega_k}{4\pi}\nonumber\\&\times&\vec{V}_{\rm F}^\eta \frac{\vec{V}_{\rm F}^\eta\cdot\frac{\partial \phi^\eta}{\partial \vec{R}}+e\vec{V}_{\rm F}^\eta \times \vec{B}^{\rm em}\cdot\frac{\partial \phi^\eta}{\partial \vec{k}_{\parallel}}}{(\omega_n^2+|\Delta|^2\sin^2 \theta^\eta)^{3/2}}|\Delta|^2,
\end{eqnarray}
where 
$\nu(0)$ is the density of state at the Fermi energy. From numerical calculations, it is found that the $z$-component of Eq.~(\ref{ssc}) is nonzero, while the $x$ and $y$ components vanish. Therefore, in this case, the emergent chiral magnetic field induces a spin supercurrent parallel to the field. 

We will demonstrate below that to obtain nonzero $\vec{j}^{\rm spin}_{\rm s}$, we need anisotropy of the Fermi surface: i.e. the Fermi velocity on a Fermi surface which surrounds a Dirac point should be anisotropic.
This anisotropy suppresses cancellation of spin supercurrents due to the average over the momentum direction in Eq.(\ref{ssc}). As a matter of fact, the Fermi surfaces of DSMs surrounding Dirac points are generally anisotropic in realistic situations, and hence, we have nonzero spin supercurrent induced by a chiral magnetic field. The anisotropy of the Fermi surface is parametrized by effective mass $M_{xx}=M_{yy}$ in our model. The  $1/M_{xx}$ term of the Hamiltonian changes the shape of Fermi surfaces from spheroid to asymmetric one (FIG.~\ref{ani}(a)-(c)). We show the anisotropy dependence of the spin supercurrent in FIG.~\ref{ani}. In this calculation, we dropped the term proportional to $1/M_{zz}$, which gives subleading corrections. It is found that the spin supercurrent is nonzero as long as $1/M_{xx} \neq 0$ (FIG.~\ref{ani}(d)). 
This feature is not seen in the case of a single band WSC model considered in Sec.\ref{sec:level2}, where anisotropy of the Fermi surface does not play a role. This is due to the oversimplification of the single band model. It is expected that for more realistic multi-orbital WSC models, anisotropy of the Fermi surface is necessary for the realization of nonzero supercurrent induced by a chiral magnetic field.

The above result also implies that, if bulk current flow is prohibited by boundary conditions of the system, the spin-dependent Fulde-Ferrell state occurs. In this Fulde-Ferrell state, the superconducting gap function $\Delta_{s_z}$ for spin $s_z=\pm 1$ pairs is modulated as,
\begin{eqnarray}
\Delta_{s_z}(z)=\Delta_0e^{is_zQz},
\end{eqnarray}
with $Q \propto B^{\rm em}$.

\begin{figure}[h]
\begin{center}
\includegraphics[width=75mm]{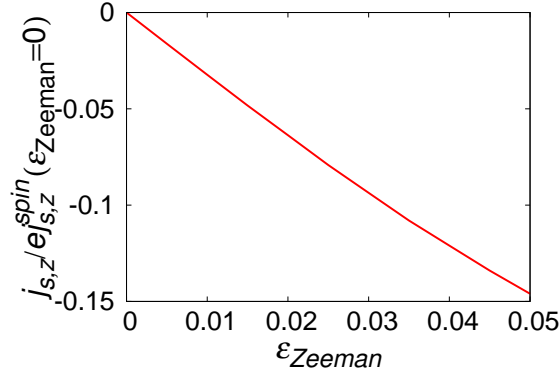}
\end{center}
\vspace{5mm}
\caption{Charge supercurrent $\vec{j}_{sz}$ divided by the electric charge $e$ plotted as a function of the Zeeman energy 
$\varepsilon_{\text{Zeeman}}$. The magnitude of the supercurrent is normalized by
the value of a spin supercurrent for $\varepsilon_{\text{Zeeman}}=0$.
The chemical potential and the effective mass is set as $\mu=0.05, M_{xx}=0.89, M_{zz}=-1.04$. The other parameters are the same as those used in FIG.\ref{ani}.
}
\label{jtz}
\end{figure}

We, furthermore, investigate effects of a Zeeman magnetic field on spin and charge supercurrents. 
The Zeeman field results in spin polarization, leading to nonzero net charge supercurrents. 
 In FIG.~\ref{jtz}, we show a calculated result for a charge supercurrent, $j_{{\rm s},z}$, induced by the uniform Zeeman field
parallel to the $z$-axis.
In this figure, the magnitude of the charge supercurrent is normalized by the magnitude of the spin supercurrent for zero magnetic field, $j^{\rm spin}_{{\rm s},z}$.

\section{\label{sec:level5}Summary}
We have investigated effects of strain-induced chiral magnetic fields on Cooper pairs in Weyl and Dirac superconductors. It is found that although the chiral magnetic field leads to neither Meissner effect nor vortex states, the pseudo-Lorentz force due to the chiral magnetic field affects dynamics of Cooper pairs drastically, and gives rise to charge/spin supercurrent in Weyl/Dirac superconductors. These effects are akin to the CME in WSMs. Accordingly, the Fulde-Ferrell state characterized by spatially inhomogeneous phase of the superconducting order parameter is realized by the chiral magnetic field, if bulk current flow is prohibited by boundary conditions. Our findings for DSCs are relevant to the DSM material Cd$_3$As$_2$ which shows superconductivity under applied pressure $\sim 8.5$ GPa.~\cite{L.P.He}  Also, WSCs can be realized in superlattice structures composed of a topological insulator and a conventional superconductor, as proposed in Ref.~\onlinecite{bale-weyl}.

Although the experimental detection of the Fulde-Ferrell state is quite difficult, supercurrents induced by strain can be observed via various methods. For instance, for a ring-shape WSC sample, strain-induced supercurrents generate a magnetic field penetrating the ring, which is experimentally measurable.  
For the case of DSCs, a spin supercurrent results in spin accumulation on surfaces of the sample, which can be also detected.~\cite{spinaccu1,spinaccu2}
Experimental exploration for these effects in real materials remains as interesting future issues.

\begin{acknowledgments}
The authors are grateful to J. de Lisle, A. Tsuruta, and Y. Yanase for useful discussions.
This work was supported by the Grant-in-Aids for Scientific
Research from MEXT of Japan [Grants No.~JP17K05517, No.~25220711, and No.~JP16K05448] and KAKENHI on Innovative Areas ``Topological Materials Science'' [No.~JP15H05852, No.~JP15H05855 and No.~JP15K21717]. 
\end{acknowledgments}

\end{document}